# Estimation of the FR4 Microwave Dielectric Properties at Cryogenic Temperature for Quantum-Chip-Interface PCBs Design

Alessandro Paghi, *Member, IEEE*, Giacomo Trupiano, Claudio Puglia, Hannah Burgaud, Giorgio De Simoni, Angelo Greco, and Francesco Giazotto

*Abstract*—Ad-hoc interface PCBs (Printed Circuit Boards) are today the standard connection between cryogenic cabling and quantum chips. Besides low-loss and low-temperature-dependent-dielectric-permittivity materials, Flame Resistance n.4 (FR4) provides a low-cost solution for fabrication of cryogenic PCBs. Here, we report on an effective way to evaluate the dielectric performance of a FR4 laminate used as substrate for cryogenic microwave PCBs. We designed a coplanar waveguide $\lambda/2$ open-circuit series resonator, and we fabricated the PCB using a low-cost manufacturing process, obtaining in-plane geometric features with maximum variations of 50-100 μm compared to the PCB design. Such a geometry allows to exploit the resonance peak of the resonator to measure the variation of the complex permittivity as a function of the temperature. The resonance peak frequency was used to estimate the real permittivity, achieving a sensitivity of -470 MHz and a resolution of $1.2 \times 10^{-2}$. Similarly, the resonance peak magnitude was involved in the extrapolation of the loss tangent, reaching a sensitivity of ~-337 dB and a resolution of $1.6 \times 10^{-4}$. For the FR4 laminate used, we estimated a 9 % reduction of the real permittivity and a 70 % reduction of the loss tangent in the temperature range from 300 to 4 K. The proposed approach can be immediately extended to the detection of cryogenic temperature-dependent dielectric performance of any kind of substrate.

## I. INTRODUCTION

IN the last years, physicists and engineers focused on the development of new materials, devices, and technologies for quantum nanoelectronics [1], [2]. Future quantum architectures will likely require the readout, manipulation, and interaction of several on-chip quantum bits, nanoelectronic devices, and control systems [3], [4]. As achieved in today's integrated circuits, it is reasonable to assume that future intra- and inter- connections in quantum architectures could be achieved by lithographically defined traces, removing the limit to directly address each individual quantum component from the outside. However, the current state of the art on quantum electronics requires the direct readout and manipulation of any aspect of the proposed device at cryogenic temperature, typically controlled by external measurement setups at room temperature [5], [6]. Ad-hoc interface printed circuit boards (PCBs) provided with DC and RF connectors, are today the standard interface between cryogenic cabling and quantum chips. Applications are reported in cryogenic filtering, microwave sample holder, and bias tee PCBs [7], [8]. Increasing the complexity of quantum chips, more sophisticated interface PCBs with a higher number of electrical connections and components are required [5]. In this framework, low-loss and low-temperature-dependent-dielectric-permittivity materials, such as those produced by the Roger Corporation (e.g., Rogers™ 4003C and Rogers™ 4350B), have been in a leading position in the developing of microwave and millimeter-wave ad-hoc cryogenic PCBs, compared to more standard substrates as the Flame Resistance n.4 (FR4) [7], [8]. Yet, manufacturing PCB prototypes with FR4 could be 10-to-50 times cheaper than the Rogers™ counterpart. It is known that FR4 dielectric properties are temperature dependent and are not usually provided in the cryogenic temperature range by the manufacturers [9]. Consequently, it is worth characterizing commercial FR4 laminates to estimate their dielectric properties to improve the PCB prototyping process at cryogenic temperatures. Transmission-reflection and resonance methods are generally employed to retrieve the complex permittivity of PCB laminates [10]–[12]. Transmission-reflection methods permit the measurement in a wider frequency range. On the other hand, resonance methods use a single frequency (or a set of frequencies), allowing the highest available accuracy for estimation of real and imaginary permittivity. Among resonance methods, resonant cavities operate with modes resonating between metallic walls, where wall losses must be considered [10], [13]. On the other hand, dielectric resonators operate with modes resonating within the sample substrate with minimal impact of metallic losses [10]. For both the resonance

This work was supported by the EU's Horizon 2020 Research and Innovation Framework Programme under Grant No. 964398 (SUPERGATE), No. 101057977 (SPECTRUM), and the PNRR MUR project PE0000023-NQSTI.

Alessandro Paghi, Giacomo Trupiano, Giorgio De Simoni, Angelo Greco, and Francrsco Giazotto are with Istituto Nanoscienze-CNR and Scuola Normale Superiore, Piazza San Silvestro 12, 56127 Pisa, Italy (email: alessandro.paghi@nano.cnr.it; giacomo.trupiano@sns.it; giorgio.desimoni@nano.cnr.it; angelo.greco@nano.cnr.it; francesco.giazotto@sns.it).

Claudio Puglia was with Istituto Nanoscienze-CNR and Scuola Normale Superiore, Piazza San Silvestro 12, 56127 Pisa, Italy, now with INFN Sezione di Pisa, Largo Bruno Pontecorvo 3, I-56127 Pisa, Italy (email: claudio.puglia@pi.infn.it).

Hannah Burgaud was with Istituto Nanoscienze-CNR and Scuola Normale Superiore, Piazza San Silvestro 12, 56127 Pisa, Italy, now with Ecole Normale Supérieure, 45 rue d'Ulm, 75005 Paris, France (email: hannah.burgaud@ens.psl.eu).

An online version of supplemental materials is available.

methods, the material under test characterization could be performed using substrate test methods, which require no or little substrate processing. Examples are, but not limited to, Hakki-Coleman dielectric resonators [14], [15], single post dielectric resonators [16], split post dielectric resonators [16]–[19], clamped stripline resonators [11], [20], and others [21]–[23]. Moreover, microwave resonant circuits can be directly manufactured on the laminate under test and used to investigate its dielectric properties, which removes limitation induced by air gap trapping, exact estimation of the laminate thickness, and specific laminate sizes [11], [24]–[26]. Table S1 summarizes the resonant methods involved in the complex permittivity estimation.

Here, we report on an effective way to evaluate the dielectric performance of a FR4 laminate used as substrate for cryogenic microwave PCBs. We designed a grounded coplanar waveguide (GCPW) $\lambda/2$ open-circuit series dielectric resonator, and we fabricated the PCB using a low-cost commercially available manufacturing process. Such a geometry allows us to exploit the resonance peak of the resonator to measure the variation of the complex permittivity as a function of the temperature. The resonance peak frequency was used to estimate the real permittivity, achieving a sensitivity of -470 MHz and a resolution of $1.2 \times 10^{-2}$, which is close to the highest values reported in the literature [13], [23]. Similarly, the resonance peak magnitude was used to extrapolate the loss tangent, reaching a sensitivity of ~-337 dB and a resolution of $1.6 \times 10^{-4}$. The manufacturing capabilities enable the fabrication from 100 µm to several cm in-plane geometric features with maximum variations of 50-100 µm compared to the PCB design. For the FR4 laminate under test, we estimated a 9 % reduction of the real permittivity and a 70 % reduction of the loss tangent in the temperature range from 300 K to 4 K, which should be considered in the cryogenic PCB design.

## II. MATERIALS AND METHODS

### A. PCB Design, Simulation, and Fabrication

The PCB was designed with KiCAD (v5.1.7) using front and bottom copper (Cu) level and manufactured exploiting the PCBWay *Normal process* [27]. FR4 (Kingboard Laminates Holdings Ltd., laminate KB-6165F, prepreg KB-6065F, 1.6 mm thickness, 150 °C glass transition temperature) was used as dielectric substrate for the PCB, with Cu layers on the top and bottom with a minimum track/spacing of 127 µm and minimum vias hole size of 300 µm (tenting vias technology). 1 oz Cu was chosen as Cu finish and Electroless Nickel Immersion Gold (ENIG) as surface finish. The PCB was simulated with Sonnet® (v18.56, cell size of 25 µm, ABS sweep from 300 kHz to 8.5 GHz, Fig. S1) carefully reproducing the design through morphological measurements taken from each PCB measured (Table S2-5). We notice that the cell size used is 5 times smaller than the minimum feature implemented in the PCB design, namely the coupling gap (G).

### B. Cryogenic Measurement System

SMA connectors (Amphenol RF, 901-10112) were soldered to the PCBs and used to test the RF electrical behavior of the device. PCBs were provided with an ad-hoc manufactured back Cu (Cu-ETP CW004A EN 13601) thermal plate and covered with an ad-hoc Al top faraday cage (38 mm × 10 mm × 3 mm). The latter was used to avoid unwanted resonances induced by the cm-size cavity created by the 40 K Cu cylindrical shield (Fig. 1a, green dashed line) of the closed cycle cryocooler (Advanced Research System, DE-210) we employed in the temperature (T) characterization of the device (Fig. 1b). PCBs were mounted in contact with the 4 K cold head (Fig. 1a, red solid line) and connected with coaxial cables through SMA connectors (Fairview Microwave, FMC0202085, Fig. 1a, blue solid line). Power attenuators (Mini-Circuits, BW-S3-2W263+, 3 dB attenuation) were used to thermalize electronic temperatures both at 40 K and 4 K. We performed temperature sweeps from 4 K to 300 K at a chamber residual pressure of $7 \times 10^{-6}$ mbar, measuring the scattering parameters in the frequency range from 300 kHz to 8.5 GHz (1.7 MHz step, 0 dBm input power) using a Vector Network Analyzer (VNA, Pico Technology, PicoVNA 108).

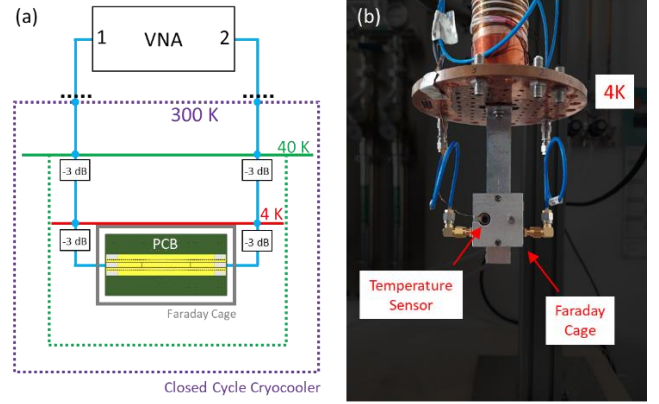

**Fig. 1.** The microwave cryogenic setup implemented. (a) Schematic diagram of the measurement setup. Violet, green, and red dashed and solid lines are used to identify thermal shields and plates, respectively. Blue solid lines identify electrical cables. Black dashed lines identify the VNA reference planes. (b) Picture of the PCB mounted on the 4K cold head and connected to the VNA using coaxial cables, highlighting the ad-hoc faraday cage manufactured and the position of the temperature sensor used.

The calibration of the VNA was performed with the short, open, load, and through procedure (SOLT, Pico Technology, Standard SOLT female SMA cal-kit). At 300 K, we measured all the two-port S-parameter magnitudes ($|S_{11}|$, $|S_{21}|$, $|S_{12}|$, and $|S_{22}|$) by putting the reference planes prior the PCB SMA connectors (*method 1*). On the other hand, for the on-cryostat measurements, we collected only $|S_{21}|$ by moving the reference planes at the input of the cryostat (Fig. 1a, black dashed lines, *method 2*) [28], [29]. The need for *method 2* for the cryogenic calibration is related to the temperature-dependent behavior of the measurement performed, since, in addition to the PCB, also coaxial cables and power attenuators show temperature-dependent electrical-performance [28]. By moving the reference planes at the input of the cryostat, the collected temperature-dependent $|S_{21}|$ include both the electrical performance of the PCB and of the connecting cables/power attenuators used. This specific $|S_{21}|$ file, obtained temperature-by-temperature, is named $|S_{21}|_{PCB+Sys}$. To extrapolate $|S_{21}|$ referring only to the PCB itself, named $|S_{21}|_{PCB}$, we replaced in a successive cooldown the measured PCB with a reference through connector (Fig. S2), collecting the temperature-dependent $|S_{21}|$. This allows us to extrapolate temperature-by-temperature the electrical performance of the cryostat electrical cabling. The $|S_{21}|$ file obtained is named $|S_{21}|_{Sys}$. Eventually, we extrapolated $|S_{21}|$ of the PCB itself solving the following equation temperature-by-temperature:

$$|S_{21}|_{PCB}=|S_{21}|_{PCB+Sys}-|S_{21}|_{Sys} \quad (1)$$

We only measured magnitude of S-parameters since phase value is not involved in extrapolation of any of the PCB electrical properties.

## III. RESULTS AND DISCUSSION

### A. PCB Design

Fig. 2 shows the CAD of the manufactured PCB.

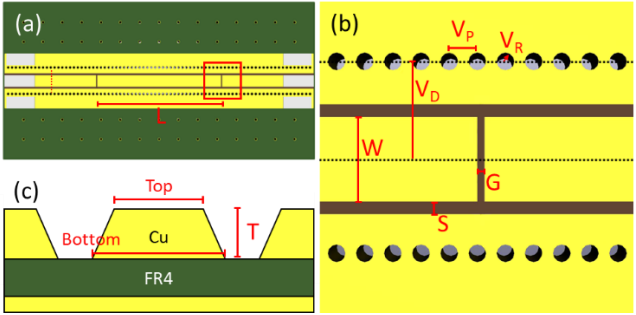

**Fig. 2.** CAD of the manufactured PCB. (a) Overall top view of the PCB exhibiting the 5 GHz resonator and GCPW. (b) Detail of the resonator capacitive coupling area in red square of (a) showing the coupling gap, the signal line-ground plane spacing, and ground vias. (c) In-section view of the red dashed lines in (a) exhibiting bottom and top Cu thicknesses.

We designed a GCPW $\lambda/2$ open-circuit series resonator (Fig. 2) employing length (L) of 16 mm and width (W) of 1.5 mm connected with a GCPW by the two coupling capacitors, one for each side, exploiting a gap (G) of 125 μm. We matched the impedance of the GCPW and the resonator to $Z_0=50$ Ω with a signal-line-ground separation (S) of 225 μm. Higher order resonance modes suppression was obtained with ground vias placing ($V_D = 1.6$ mm $< \lambda_{MAX}/4$, $V_P = 0.6$ mm $< \lambda_{MAX}/4$, $V_R = 0.15$ mm, with $\lambda_{MAX} = c/(f_{MAX}\times\varepsilon_{R\,FR4})$, $c = 3\times10^8$ m/s, $f_{MAX} = 8.5$ GHz, $\varepsilon_{R\,FR4} = 4.6$) [30]. PCB design values are summarized in Table 1.

**TABLE I**
*PCB DESIGN PARAMETERS*

| Design Parameter | Design Value [μm] |
|---|---|
| L | 16000 |
| W | 1500 |
| S | 225 |
| G | 125 |
| $V_D$ | 1600 |
| $V_P$ | 600 |
| $V_R$ | 150 |
| t | 35 |

Fig. 2c shows the in-section view of the PCB depicting the typical Cu layer trapezoidal profile obtained via Cu acidic wet etching involved in trace definition manufacturing step. The trapezoidal profile allows to define each in-plane geometry both on top and on bottom of the Cu trace.

### B. Morphological Characterization of the Manufactured PCB

Fig. 3a shows a typical scanning electron microscopy (SEM, Zeiss, Merlin Gemini, 5 kV acceleration voltage, 30× magnification, 34° tilt, secondary electron detector) image of the coupling gap depicted in Fig. 2b.

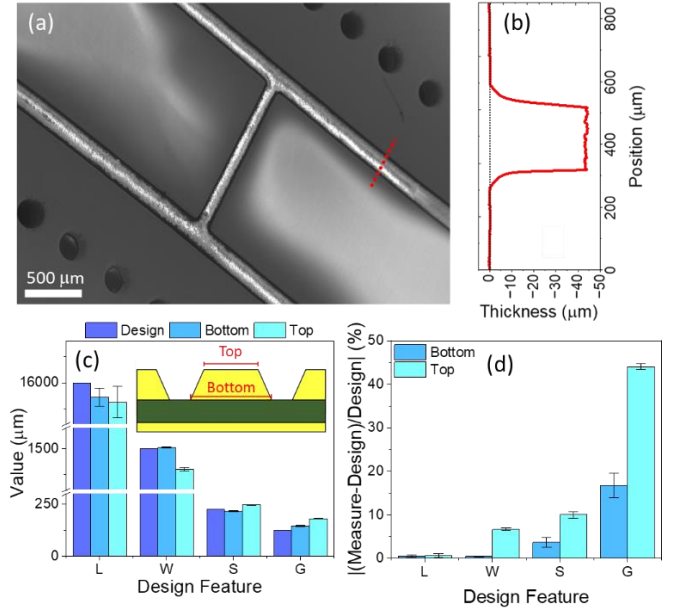

**Fig. 3.** Morphological characterization of the manufactured PCB. (a) SEM image of the coupling gap area. (b) Stylus profilometer analysis performed in the red dashed line of (a) showing Cu trace thickness. (c) Designed and measured top and bottom values of in-plane PCB geometries. The inset depicts the in-section PCB view. (d) Absolute normalized relative variation of measured vs. designed in-plane PCB geometries. Data in (c,d) are reported as the average value measured over 3 PCBs, with error bars representing the standard deviation.

The micrograph shows that the PCBWay *Normal Process* involved in the PCB manufacturing allows to achieve in-plane features in the range from 100 μm to several cm. A clear separation of the Cu traces (dark grey area) is observed from the FR4 substrate left exposed on the bottom side (light gray area). Cu traces are about 43 μm thick (Fig. 3b) (Bruker, DektacXT, stylus radius 12.5 μm, force 3 mg).

Fig. 3c shows the comparison between the design parameters (dark blue) and top/bottom measured dimensions (light blue). We measured a maximum variation of about 100 μm in the case of L (bottom) value, compared to the design of 16 mm. Regarding G, we found that the fabrication process led to a top maximum value of 55 μm larger than the designed (125 μm). Fig. 3d displays the absolute normalized relative variations (ANRV), defined as:

|Measured value - Design value|/Design value     (2)

for both top and bottom in-plane measured geometries. These results point out that for smaller features the fabrication error leads to larger ANRV. Specifically, we found a maximum 44 % variation for G (top), the smallest feature of the PCB. Such values are usually better than the in-plane fabrication tolerance of ≥ 20 % reported for the PCBWay Normal process [27].

*C. Microwave Electrical Behavior of the PCB at 300 K*

Fig. 4 shows typical measured and simulated S-parameters of the PCB at the temperature of 300 K.

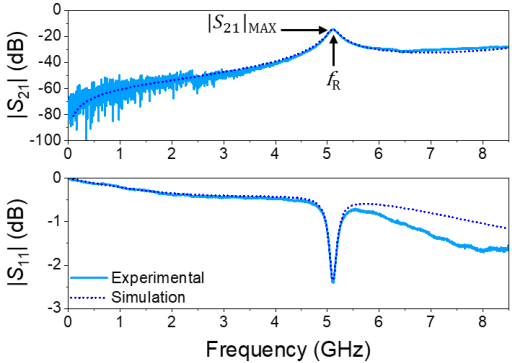

**Fig. 4.** Typical measured and simulated S-parameters of the PCB at 300 K.

A resonance peak (dip) is evident in $|S_{21}|$ ($|S_{11}|$) from which values of resonance frequency ($f_R$) and $|S_{21}|$ magnitude ($|S_{21}|_{MAX}$) were extrapolated. We also calculated the quality factor ($Q_{-10dB}$) as $Q_{-10dB}=f_R/BW$, where $BW$ is the frequency bandwidth where the resonance peak value is reduced by -10 dB. We obtained values of $f_R$ = 5.104 ± 0.011 GHz, $|S_{21}|_{MAX}$ = -14.390 ± 0.304 dB, and $Q_{-10dB}$ = 9.27 ± 0.23 (avg and std evaluated with 3 PCBs). We extrapolated the coefficient of variation (CV% = 100 × average value / std) of $f_R$, $|S_{21}|_{MAX}$, and $Q_{-10dB}$, which correspond respectively to 0.2%, 2% and 2.5%. We note that the CVs just calculated are very similar to the CVs of the geometrical features, hence 0.3% for L and 1.7% for G. This is not surprising since L and G are the geometrical features mostly affecting $f_R$ and $Q_{-10dB}$, respectively.

The simulation was carried out in Sonnet® with the design reported in Fig. S1 using the measured geometrical parameters, which are reported in Table S2-5. In the following, we will refer to the complex permittivity as $\varepsilon^*=\varepsilon_R-j\varepsilon_I$, where $\varepsilon_R$ and $\varepsilon_I$ are real and imaginary permittivities (expressed relative to the free-space), respectively. The parameters extracted from the best-fitting procedure of the peak resonance frequency and magnitude (Fig. 4) are $\varepsilon_R$ and the loss tangent ($tan\delta = \varepsilon_I/\varepsilon_R$). At 300 K, we extrapolated $\varepsilon_R$ = 4.80 ± 0.04 (avg and std evaluated with 3 PCBs), in agreement with specification reported by the PCB laminate manufacturer, namely $\varepsilon_R$ ≤ 5.4. In a similar way, we retrieved $tan\delta$ = 0.033 ± 0.003 (avg and std evaluated with 3 PCBs), which is also according with $tan\delta$ ≤ 0.035 reported as specification value. Results were validated through more simulations performed with COMSOL Multiphyisics®, observing an excellent agreement of both the simulation methods involved and experimental results (Fig. S6).

Eventually, we evaluated the effect of the process variance on the $f_R$ and $|S_{21}|_{MAX}$ values, with simulation results shown in Fig. S3. We found that L is the most critical parameter to take into account for $f_R$, with a variation of 90 MHz in the ± 3σ range. On the other hand, the G value is the most relevant parameter for $|S_{21}|_{MAX}$, which induces a variation of about 2 dB in the ± 3σ range.

*D. Temperature-Dependent Microwave Electrical Behavior of the PCB*

We then measured the temperature-dependent microwave electrical properties of the fabricated PCB with the measurement setup reported in Fig. 1, extrapolating the dielectric properties vs. temperature behavior of the FR4 substrate chosen. To retrieve the impact of both $\varepsilon_R$ and $tan\delta$ variations on the electrical behavior of the microwave resonator embedded on the PCB, we systematically changed $\varepsilon_R$ and $tan\delta$ one value at a time in Sonnet®, starting from the best-fitting values obtained at 300 K. We collected $f_R$ and $|S_{21}|_{MAX}$ from simulated spectra as a function of $\varepsilon_R$ and $tan\delta$ and we best-fitted the experimental trends with theoretical curves. Fig. 5 shows typical sensing curves extrapolated from simulated spectra, where dots represent values obtained from simulations while solid lines the best-fitting curves, depicting in Fig. 5a the $f_R$ vs. $\varepsilon_R$ and $tan\delta$ behaviors, and in Fig. 5b the $|S_{21}|_{MAX}$ vs. $\varepsilon_R$ and $tan\delta$ trends.

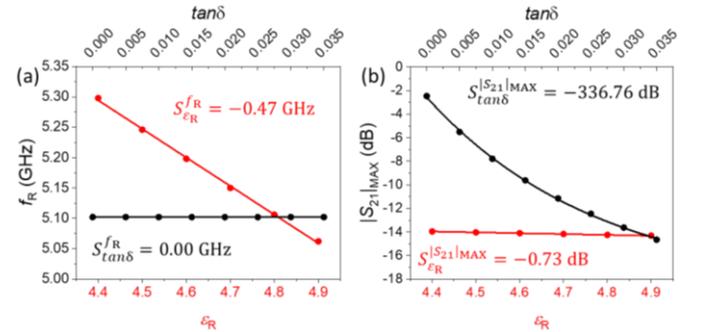

**Fig. 5.** Typical sensing curves extrapolated from simulated spectra. (a) Resonance frequency vs. real permittivity (down) and loss tangent (up). (b) $|S_{21}|$ magnitude vs. real permittivity (down) and loss tangent (up).

The theoretical curves used are linear, in the case of $f_R$ vs. $\varepsilon_R$, $f_R$ vs. $tan\delta$, and $|S_{21}|_{MAX}$ vs. $\varepsilon_R$:

$$f_R = m_1 \times \varepsilon_R + q_1 \quad (3)$$
$$f_R = m_2 \times tan\delta + q_2 \quad (4)$$
$$|S_{21}|_{MAX} = m_3 \times \varepsilon_R + q_3 \quad (5)$$

where $m_i$ is the angular coefficient and $q_i$ is the $f_R$- or $|S_{21}|_{MAX}$-axis intercept, and a first order exponential decay for $|S_{21}|_{MAX}$ vs. $tan\delta$:

$$|S_{21}|_{MAX} = a_4 \times exp(-tan\delta / t_4) + q_4 \quad (6)$$

where $a_4$ is the exponential starting value, $t_4$ is the exponential time constant, and $q_4$ is the $|S_{21}|_{MAX}$-axis intercept. Values of fitting parameters are reported in Table S6.

An excellent agreement between best-fittings and data retrieved from simulations was achieved. We defined $S_B^A$ as the sensitivity of $A$ to variations of $B$, which is evaluated as the angular coefficient of the linear best fitting of the $A$ vs. $B$ curve; from the best fitting curves, we extrapolated $S_{\varepsilon_R}^{f_R} = -0.47$ GHz, $S_{tan\delta}^{f_R} = 0.00$ GHz, $S_{tan\delta}^{|S_{21}|_{MAX}} = -336.76$ dB, $S_{\varepsilon_R}^{|S_{21}|_{MAX}} = -0.73$ dB, and we reported these values in Fig. 5. We notice that we find negligible impact of variation of $tan\delta$ on the $f_R$ value and $\varepsilon_R$ on the $|S_{21}|_{MAX}$ value, which allow us to use $f_R$ as a selective sensing parameter for $\varepsilon_R$ and $|S_{21}|_{MAX}$ as a selective sensing parameter for $tan\delta$, under the approximation of negligible changes of in-plane and out-of-plane PCB geometric features with the temperature [9]. Remarkably, the calculated sensitivities are close to the highest values reported in the literature [13], [23].

Fig. 6a shows the typical measured $|S_{21}|$ spectra at different temperatures, from which both an increase of $f_R$ and $|S_{21}|_{MAX}$ are clearly observed decreasing the temperature.

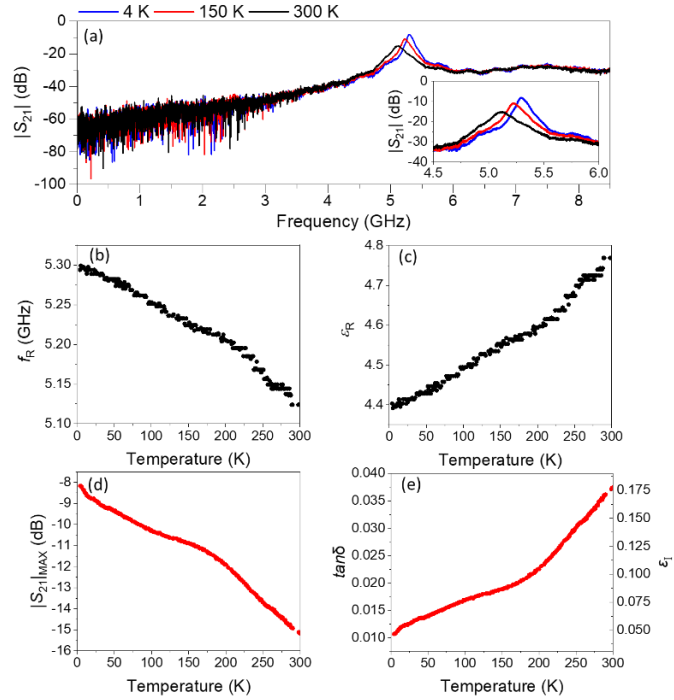

**Fig. 6.** Typical temperature-dependent microwave electrical behavior of the PCB and FR4 laminate dielectric properties. (a) Measured temperature-dependent $|S_{21}|$ spectra. The inset shows the resonance peaks. (b) Measured resonance frequency vs. temperature. (c) Extrapolated real permittivity vs. temperature. (d) Measured $|S_{21}|$ maximum vs. temperature. (e) Extrapolated loss tangent and imaginary permittivity vs. temperature.

Fig. 6b shows the $f_R$ vs. temperature trend; a monotonic increase of the resonance frequency from about 5.1 to 5.3 GHz was collected by decreasing the temperature. We extrapolated the $\varepsilon_R$ vs. temperature behavior inverting equation (3) in:

$$\varepsilon_R = (f_R(T) - q_1)/m_1 \quad (7)$$

and substituting $f_R$ values extrapolated from the temperature-dependent $|S_{21}|$ spectra to achieve the experimental $\varepsilon_R$ vs. temperature curve. Fig. 6c shows values of the permittivity in the temperature range 4 K to 300 K, where a monotonic decrease from $\varepsilon_R = 4.82 \pm 0.05$ @ 300 K to $\varepsilon_R = 4.40 \pm 0.05$ @ 4 K is calculated (avg and std evaluated with 3 PCBs).

The $|S_{21}|_{MAX}$ vs. temperature behavior is shown in Fig. 6d, while the $Q_{-10dB}$ vs. temperature trend was reported for completeness in Fig. S4. As reported for the resonance frequency, also in this case a monotonic increase of the maximum of the $|S_{21}|$ spectrum from about -15.5 to -8 dB was observed by decreasing the temperature. As previously reported, we retrieved the $tan\delta$ vs. temperature behavior, as well as the $\varepsilon_I$ vs. temperature relationship, inverting equation (6) in:

$$tan\delta = -t_4 \times ln(|S_{21}|_{MAX}(T) - q_4)/a_4 \quad (8)$$

and substituting $|S_{21}|_{MAX}$ values collected from the temperature-dependent $|S_{21}|$ spectra to obtain the experimental $tan\delta$ vs. temperature trend. Fig. 6e shows both values of the loss tangent and the imaginary permittivity in the temperature range 4 to 300 K, where a monotonic decrease from $tan\delta = 0.038 \pm 0.004$ @ 300 K to $tan\delta = 0.012 \pm 0.002$ @ 4 K is calculated (avg and std evaluated with 3 PCBs). We notice that the behavior observed for both $\varepsilon_R$ and $tan\delta$ is in agreement with what reported in the literature [9], [31], although a microscopic explanation of why this happens for the FR4 is not given to date. More information about the electrical field intensity and direction for the modelled PCB are reported in the Supporting Information section "COMSOL Multiphyisics® Model Implementation and Simulation". More information about the evaluation of the effect of the dielectric permittivity anisotropy on $f_R$ and $|S_{21}|_{MAX}$ are reported in the Supporting Information section "Evaluation of the effect of the dielectric permittivity anisotropy on $f_R$ and $|S_{21}|_{MAX}$".

We defined the resolution as the lowest variation of the target quantity, i.e., $\varepsilon_R$ or $tan\delta$, that produces a change of the sensor output, namely, $f_R$ or $|S_{21}|_{MAX}$, that can be discriminated from the noise floor with a stated confidence of 3 times the standard deviation of the sensor output $O$. We found a resolution of 5.7 MHz for $f_R$ and 0.051 dB for $|S_{21}|_{MAX}$ that are traduced in a resolution of $1.6 \times 10^{-2}$ for $\varepsilon_R$ and $1.6 \times 10^{-4}$ for $tan\delta$, which are considered enough to carefully estimate the dielectric properties of a substrate laminate for design of PCBs at cryogenic temperatures.

*E. Application to a Specific Case*

Eventually, we focused on the simulation of the microwave performance of a cryogenic PCB embedding a GCPW at 4 K compared to 300 K (Fig. 7a). GCPWs are usually employed as communication links on quantum chips and interface PCBs to transmit microwave signals. The designed GCPW employs a length of 38 mm and a width of 1.5 mm. Signal-line-ground separations of 250 and 300 μm were used to match the characteristic impedance of the GCPW to $Z_0$=50 Ω at 4 and 300 K, respectively.

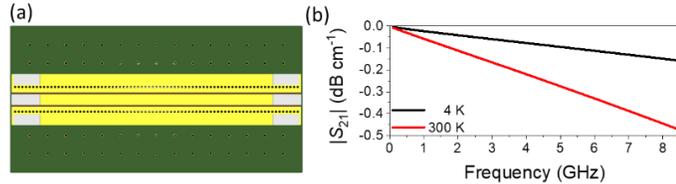

**Fig. 7.** Impact of the permittivity on the design of a cryogenic microwave PCB. (a) CAD of an on-PCB GCPW. (b) Simulated insertion losses of an on-PCB GCPW at 4 K and 300 K.

As shown in Fig. 7b, a ~66 % reduction of the insertion losses ($|S_{21}|$, from -0.275 dB cm$^{-1}$ at 300 K to -0.095 dB cm$^{-1}$ at 4 K evaluated at 5 GHz) is achieved at 4 K compared to 300 K, which is related to the reduction of the loss tangent as the device temperature decreases.

We note that low-loss permittivity materials generally employed in cryogenic interface PCBs, like those produced by Rogers™, present $tan\delta$ one order of magnitude lower than the FR4 estimated here [18]. Nonetheless, the standard selling price for the formers is one order of magnitude higher than the latter. This fact imposes an important trade-off to take into account when engineering cryogenic PCBs.

## IV. CONCLUSIONS

In this work, we reported an effective procedure to evaluate the dielectric performance of a commercial FR4 laminate used as substrate for cryogenic microwave PCBs using a GCPW $\lambda$/2 open-circuit series resonator. We fabricated the PCB using a low-cost manufacturing process, obtaining in-plane geometric features with maximum variations of 50-100 μm compared to the PCB design. We used the resonance peak frequency and magnitude as sensing parameters for the real permittivity and loss tangent, estimating a reduction of about 9 and 70 % from 300 to 4 K, respectively. The temperature-dependent dielectric permittivity values were then used to design a cryogenic-specific FR4-based PCB embedding a GCPW. Such a cryogenic specific design showed a reduction of ~66 % of the insertion losses at 4 K, compared to the results achieved at 300 K.

## APPENDIX

*A. Error on $|S_{21}|$ spectra measurement*

We used the $|S_{21}|$ spectra measured at 300 K with *method 1* as gold standard for estimation of the microwave behavior of the PCB, since the reference planes of the VNA are directly placed at the input of the PCB SMA connectors with no spectra subtraction required. Compared to the spectra taken with method 1, at 300 K with *method 2* we estimated an error of ~-1 dB on $|S_{21}|_{MAX}$, while the error made on the evaluation of $f_R$ is similar to the frequency sampling step. This leads to an error of ~14 % and <1 % in estimation of $tan\delta$ and $\varepsilon_R$, respectively.

*B. Supporting Information*

Supporting information can be downloaded at the link: https://doi.org/10.1109/TIM.2024.3372217/mm1

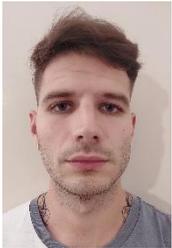

**Alessandro Paghi** (Member, IEEE) was born in Arezzo (AR), Italy in 1994. He received his B.S. (cum laude) in Information Engineering in 2016 from the University of Siena, his M.S. and Ph.D. (both cum laude) in Electronic and Information Engineering from the University of Pisa in 2018 and 2022, respectively.

Until November 2022, he was a postdoctoral researcher at the Information Engineering Department of the University of Pisa. Currently, he is a postdoctoral researcher at the Superconducting Quantum Electronics Lab (SQEL) at the Nanoscience Institute of the Italian National Research Council (CNR, Pisa) focusing on the engineering of the electrostatic field-effect in all-metallic or metallic-semiconductive-metallic superconducting devices. His main research interests focused on the development of micro and nano -structured materials, devices, and systems for unconventional DC-to-GHz electronics and UV–VIS–NIR photonics. He published 14 articles, 7 of these published as first author contributor. In 2021, he won the Huawei Italian University Challenge.

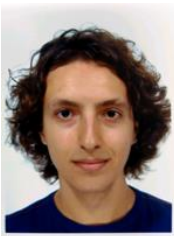

**Giacomo Trupiano** was born in Palermo (PA), Italy in 1998. He received his B.S, M.S degrees in physics from the University of Pisa in 2020 and 2022, respectively.

Since 2022 he has been a PhD student in Nanoscience at the Scuola Normale Superiore (SNS, Pisa). He is currently studying the electric field effect in all-metallic superconductive nanodevices at the Superconducting Quantum Electronics Lab (SQEL) of the Nanoscience Department of the Italian National Research Council (CNR, Pisa).

Among his research interests are superconducting electronics, nanofabrication, and cryogenics.

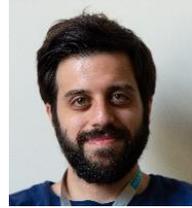

**Claudio Puglia** was born in Erice (TP), Italy in 1993. He received his B.S, M.S and Ph.D. degrees in physics from the University of Pisa in 2021.

From 2021 to 2023 he was a Researcher in the Superconducting Quantum Electronics Lab (SQEL) at the Nanoscience Institute of the Italian National Research Council (CNR, Pisa). Since 2023 he has been a Technologist at the Italian National Institute for Nuclear Physics of Pisa. He is the author of 12 articles, and 2 inventions. His research interests include superconducting electronics, cryogenics, and nanofabrication.

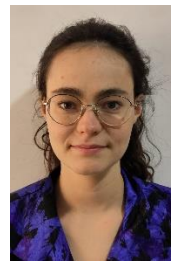

**Hannah Burgaud** was born in Nantes, France in 2000. She received her B.S. in fundamental physics from the Ecole Normale Supérieure of Paris in 2022. Since September 2022 she is doing her M.S. in fundamental physics at the Ecole Normale Supérieure of Paris and spent four months in Scuola Normale Superiore of Pisa for a research internship in the Superconducting Quantum Electronics Lab (SQEL). Her research interests are condensed matter, superconductivity, nanofluidics, energy storage and production.

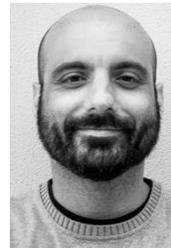

**Giorgio De Simoni** was born in Pisa (PI), Italy in 1981. He graduated in Physics (cum laude) at University of Pisa and got a Ph.D. in Condensed Matter Physics from Scuola Normale Superiore (SNS, Pisa).

After post-doc experiences at SNS, at the Italian Institute of Technology (IIT, Pisa), and at the Italian National Research Council (CNR, Pisa), he became (2016) a staff researcher of the Superconducting Quantum Electronic Lab (SQEL) of the Nanoscience Institute of CNR. His research activity focused on reduced-dimensionality quantum optoelectronics, micro-/nano-electromechanical systems, and quantum sensing. He is currently mainly active in mesoscopic superconductivity, and in superconducting electronics. He authored 4 patents and more than 30 articles with 900+ citations, reaching an h-index of 18 (according to Google Scholar).

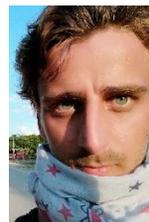

**Angelo Greco** was born in Imperia (IM), Italy in 1992. He received his M.S. in Physics of Advanced Technologies in 2018 and his Ph.D. in Metrology in 2022 (both cum laude) at the Politecnico di Torino.

He is currently working as fellow at the Nanoscience Institute of the Italian National Research Council (CNR, Pisa). His main research interests focus on superconducting quantum devices working in the microwave regime.

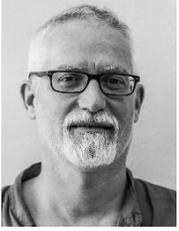 **Francesco Giazotto** was born in Rome (RM), Italy, in 1969. He graduated in Physics and got a Ph.D. in Physics (cum laude) in 2002 at Scuola Normale Superiore (SNS, Pisa). He is a research director and Principal Investigator (PI) of the Superconducting Quantum Electronics Lab (SQEL) at the Nanoscience Institute of the Italian National Research Council (CNR, Pisa). He was a visiting scientist for various periods from 2003 to 2008 at Aalto University in Helsinki (FI), and in 2011 at University Joseph Fourier in Grenoble (FR). F. Giazotto coordinates as PI the activities of mesoscopic superconductivity, coherent caloritronics, electronic refrigeration, ultrasensitive quantum magnetometry, superconducting spintronics, and quantum transport in hybrid systems at ultralow temperatures at the NEST laboratory. He has co-authored 226 articles in international journals, holds 13 patents on superconducting nanodevices, and has given more than 120 invited talks at national and international conferences. His papers have attracted more than 8050 citations and a 5-years h-index of 45 (according to Google Scholar). For his research activities in the field of thermal transport at the nanoscale he has achieved an ERC Consolidator Grant in 2013, and an ERC PoC Grant in 2020.